# INNER WARM DISK OF ESO Hα 279A REVEALED BY NA I AND CO OVERTONE EMISSION LINES

A-Ran Lyo[1], Jongsoo Kim[1,2], Jae-Joon Lee[1], Kyoung-Hee Kim[1], Jihyun Kang[1], Do-Young Byun[1], Gregory Mace[3], Kimberly R. Sokal[3], Chan Park[1], Moo-Young [1], Heeyoung Oh[1], Young Sam Yu[1], Jae Sok Oh[1], Ueejeong Jeong[1], Hwihyun Kim[4], Soojong Pak[5], Narae Hwang[1], Byeong-Gon Park[1], Sungho Lee[1], Kyle Kaplan[3], Hye-In Lee[5], Huynh Anh Nguyen Le[5], and Daniel Jaffe[3]

[1]Korea Astronomy and Space Science Institute, 776, Daedeokdae-ro, Yuseong-gu, Daejeon 34055, Republic of Korea [2]Korea University of Science and Technology, 217 Gajeong-ro, Yuseong-gu, Daejeon 34113, Republic Korea [3]Department of Astronomy, University of Texas at Austin, Austin, TX 78712, USA [4]Gemini Observatory, Southern Operations Center, c/o AURA, Casilla 603, La Serena, Chile [5]School of Space Research and Institute of Natural Sciences, Kyung Hee University, 1732 Deogyeong-daero, Giheung-gu, Yongin-si, Gyeonggi-do 17104, Republic of Korea

## ABSTRACT

We present analysis of near-infrared, high-resolution spectroscopy towards the Flat-spectrum YSO (Young Stellar Object) ESO Hα 279a (~1.5M$_\odot$) in the Serpens star forming region, at the distance of 429 pc. Using the Immersion GRating INfrared Spectrometer (IGRINS, R ≈45,000), we detect emission lines originating from the accretion channel flow, jet, and inner disk. Specifically, we identify hydrogen Brackett series recombination, [Fe II], [Fe III], [Fe IV], Ca I, Na I, $H_2$, $H_2O$ and CO overtone emission lines. By modeling five bands of CO overtone emission lines, and the symmetric double- peaked line profile for Na I emission lines, we find that ESO Hα 279a has an actively accreting Keplerian disk. From our Keplerian disk model, we find that Na I emission lines originate between 0.04 AU and 1.00 AU, while CO overtone emission lines are from the outer part of disk, in the range between 0.22 AU and 3.00 AU. It reveals that the neutral atomic Na gas is a good tracer of the innermost region of the actively accreting disk. We derive a mass accretion rate of 2 ~ 10 × $10^{-7}$ M$_\odot$ yr$^{-1}$ from the measured Brγ emission luminosity of 1.78(±0.31)×$10^{31}$ erg s$^{-1}$.

## 1. INTRODUCTION

The inner disk of young stellar objects (YSO) is crucial to terrestrial planet formation and is also the site of hot Jupiters. However, the deficiency of gas in the inner disk makes it unlikely that massive planets could have formed there. Instead, Lin et al. (1996) argue that the inner disk plays a main role in halting the migration of exoplanets and establishing their stable orbits. It is, therefore, important to identify tracers of the inner disk and to investigate the temperature and density distributions, and to study its dynamics of the inner disk to understand how it works to stop the inward migration of planets and eventually helps planets to be survived there. The inner disk is also closely related to accretion and ejection processes, which affect the formation and evolution of the central star. In a magnetospheric accretion model, material close to the corotation radius, 0.03 ~0.08 AU, accretes to a central star following magnetic field lines bridging the star and the disk (K¨onigl 1991; Shu et al. 1994; Calvet & Hartmann 1992). In addition, the predicted launching place of a jet is either at the truncation inner disk radius or in the range of 0.5 and 5 AU disk radius based on the magnetocentrifugal wind models (Shu et al. 2000; K¨onigl & Pudritz 2000). With IGRINS, we are able to study all of these factors in a single spectrum.

A high-resolution, near-infrared, spectroscopic study of inner gaseous disks is one of the best methods to derive the properties of the inner disks at 1 AU. Ro-vibrational transitions of CO molecules are good tracers of the warm (1500 ~ 5000 K) and high density (>$10^9$ cm$^{-3}$) region of disks (Scoville et al. 1983; Carr 1989; Calvet et al. 1991; Najita et al. 2000, 2003, 2007). In particular, broad CO overtone emission profiles suggest that CO gas is found close to the central star at the range of ~0.04 AU to ~0.3 AU in low mass stars (Chandler et al. 1993; Najita et al. 2000, 2003) and further out, at the range of ~0.1 AU to ~5.0 AU, in both intermediate-mass Herbig Ae/Be stars and massive YSOs (Cowley et al. 2012; Ilee et al. 2014; Wheelwright et al. 2010).

$H_2O$ emission lines at ~2.3 μm are also good tracers of the inner disk where the temperature falls in the range of 100 ~ 2500 K. Narrower line widths and a lower excitation temperature for $H_2O$ gas tell us that $H_2O$ emission comes from relatively further out in the disk than where the CO emission arises. These warm water emission lines have only been detected in a few YSOs, such as SVS 13 (Carr et al. 2004; Najita et al. 2000), DG Tau (Najita et al. 2000), V1331 Cyg (Najita et al. 2009), and IRAS 08576-4334 (Thi & Bik 2005).

In addition to CO and $H_2O$ molecules, the neutral atomic gases of Na and Ca are found in the hot inner disk regions. For example, Na I emission lines at 2.206 and 2.209 μm in the near-infrared wavelength have been detected towards Class I/Flat-spectrum YSOs HH100IR (TS 2.6), RCrA IRS2 (TS 13.1) (Nisini et al. 2005; Doppmann et al. 2005), ESO Hα 279a (Aspin et al. 1994; Aspin & Greene 2007), ISO 159, SVS 20N (Eiroa & Djupvik 2008), and IRAS 03220+3035N (Connelley & Greene 2010). Ca I emission lines at 1.978 and 1.987 μm have also been detected towards the embedded protostars and suggested to be a disk tracer (e.g., SVS 13, HH 26-IRS, HH 34-IRS, HH 72-IRS, HH 300-IRS (IRAS 04239+2436), HH 999-IRS (IRAS 06047-1117), and EX Lup; Davis et al. 2011; Kóspál et al. 2011). Yet, none of these previous observations have been studied their kinematics in detail, primarily due to the lower spectral resolution of those studies, R(= $\lambda/\Delta\lambda$) = 1,200 ~ 18,000.

ESO Hα 279a is a YSO in the Serpens molecular cloud region at the distance of 429 pc (Dzib et al. 2010, 2011). We classified it as a Flat-spectrum YSO based on the extinction corrected spectral index of -0.06 in the range of 2 μm – 24 μm provided by Dunham et al. (2015). This object is the source of the Herbig-Haro objects HH106 and HH107 that have a separation of 1.75 pc (Reipurth & Eiroa 1992; Aspin et al. 1994). Despite being found a distance of 5' away, ESO Hα 279a shows a clear connection with the main cloud based on its consistent systemic velocity of $V_{LSR}$ ~ 6.9 – 8.8 km s$^{-1}$, measured from C$^{18}$O (1-0) and $N_2H^+$ (1-0) molecular line observations (McMullin et al. 2000; Lee et al. 2014).

With velocity resolution spectra of ~15 km s$^{-1}$, Aspin & Greene (2007) showed the CO overtone emission and the double-peaked Na I emission lines in the spectrum of ESO Hα 279a. Based on the sharp uprising shape of the features and a small broadening in the blue shifted part of CO bandhead at 2.3 μm, they interpreted the emission lines originating from a wind or an accretion funnel flow rather than a circumstellar disk. They also suggested that the small broadened part of the bandhead of the first CO overtone lines, called as a shoulder (e.g. Najita et al. 1996), is not due to the broadening by disk kinematics, but due to the contribution of the $H_2O$ emission lines at 2.293 μm. However, the broad shoulder of CO bandhead is not the crucial factor to determine whether CO emissions are originated from a wind or the disk. For example, disk models work very well for the CO overtone emission lines of some YSOs lacking broad shoulders, such as MWC 349 (Kraus et al. 2000), IRAS 08576-4334, IRAS 11097-

6102 (Bik & Thi 2004), SVS 13 (Carr et al. 2004), and V1331 Cyg (Najita et al. 2009). It is the symmetric double-peaked line profile that is the crucial evidence of a Keplerian rotating disk. Therefore, the shape of resolved line profiles are necessary for judging the origin of the CO emission. In this paper, we use the advantage of high spectral resolution to examine the individually resolved R-branch lines of CO (v = 2 – 0) in the range of 2.30 – 2.32 μm. We fit them with the symmetric double-peaked line profiles, which are the characteristic ones of a Keplerian disk.

We present observation of ESO Hα 279a using IGRINS, which provides a velocity resolution of ~7 km s$^{-1}$ at 2 μm. Especially, this velocity resolution provides us to be able to study the kinematics of the inner disk by resolving disk tracer atomic and molecular emission lines of Ca I emission lines at 1.978 and 1.987 μm, Na I emission lines at 2.206 and 2.209 μm, and CO overtone emission lines in the range of 2.29~2.45 μm.

In Section 2, we describe the IGRINS observation and data reduction. We also briefly address of the identification of stellar absorption and ionized iron forbidden lines and the basic properties of the central star and jet based on the kinematical information of these spectral lines. In addition, we examine the molecular hydrogen emission line as a possible disk tracer in this source. Sections 3 presents our estimates of the mass accretion rate towards the central star using the Brγ emission line. In Sections 4 and 5, we show that disk models fit the overall line profiles of CO and Na I/Ca I emission very well; then using the best fit disk models, we drive the inner disk properties of ESO Hα 279a. In Section 6, we discuss the strengths and weaknesses of our CO model, as well as different interpretations of the symmetric double-peaked line profile of Na I. We finally summarize our results of the inner warm disk properties of the Flat-spectrum YSO ESO Hα 279a in Section 7.

## 2. OBSERVATION AND DATA REDUCTION

Near-infrared H and K-band spectra of ESO Hα 279a (R.A. = 18:29:31.97. Dec. = 01:18:42. 6 (J2000)) were obtained using IGRINS mounted on the 2.7 m Harlan J. Smith Telescope of the McDonald Observatory (see Fig. 1 and 2). IGRINS has two separate 2048×2048 pixel Teledyne HAWAII-2RG detectors that simultaneously obtain H and K-band spectra. The slit size is 1"×15" and the plate scale is 0.27" pixel$^{-1}$. IGRINS H and K-band spectra cover the entire wavelength range from 1.50 to 2.45 μm with a resolving power of R=45,000 with ~3.5 pixel sampling, corresponding to the velocity resolution of 7 km s$^{-1}$ at 2 μm (Park et al. 2014).

The observation of ESO Hα 279a was carried out during IGRINS commissioning on May 26, 2014 (UT 2014-05-27). We employed a nod-on-slit observational mode in an ABBA sequence, such that the telescope is nodded 7" along the slit in the east-west direction, to remove background sky emission from AB subtracted pairs. Total exposure time was 4×300 seconds, yielding a continuum signal-to-noise (S/N) ratio of ~120 at 2.2 μm. The spectra were calibrated using flat frames of a halogen lamp, and the wavelength solution was derived from Th-Ar arc lamp spectrum with a Python-based data reduction pipeline package (Lee 2015). V1431 Aql, a standard star with a spectral type A0V, was observed at the same air mass as ESO Hα 279a to correct for telluric absorption lines. We removed the telluric lines from the spectrum of our target source by dividing with the spectrum of the A0 standard star. We eliminated strong intrinsic photospheric hydrogen absorption lines from the standard spectrum, which results in a blackbody spectrum with telluric lines. The followings are a detailed procedure on how we have gotten the standard star spectrum with only telluric lines. (1) We made a spectrum by cutting the hydrogen absorption lines in the standard star spectrum. (2) We made another spectrum by cutting telluric lines only in the range of hydrogen absorption lines from the original standard star spectrum. (3) We divided the original standard star spectrum by the one made in (2). (4) We finally multiply the one made in (1) by the one made in (3). In addition, we manually removed some of remained abnormal spike features of the telluric lines in the edge of some of orders. Overall, the robustness of our telluric correction is seen in the range of CO overtone emission lines where most of CO R and P-branch lines have a good fitting with model (Fig 2 and see section 4).

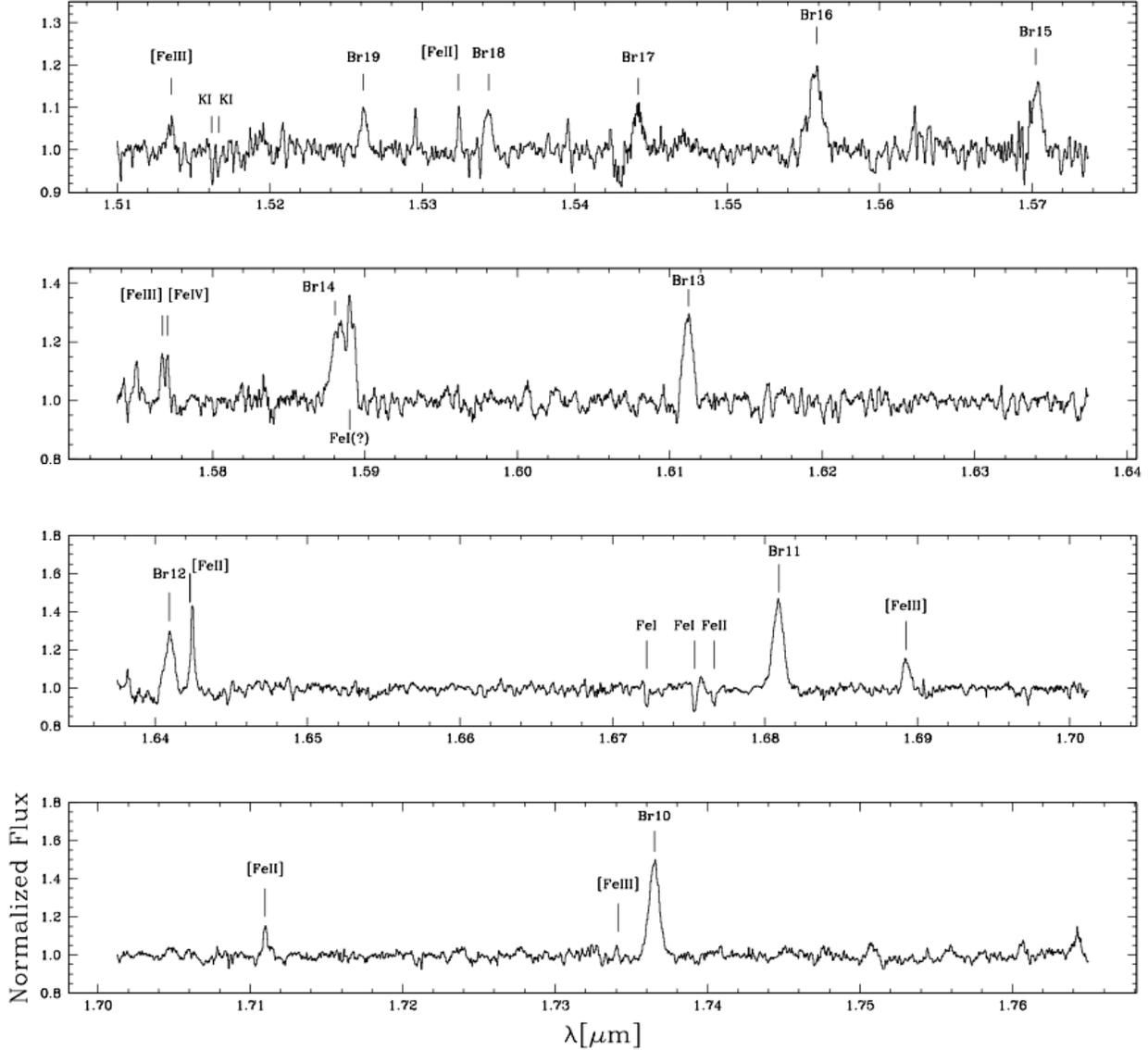

Figure 1. Telluric-corrected and normalized IGRINS spectrum of the Flat-spectrum ESO Hα 279a covering the full H-band from 1.510 μm to 1.765 μm. The identified lines are also listed in Table 1. We adopt a normalized flux level of $6.40 \times 10^{-10}$ erg s$^{-1}$ cm$^{-2}$ μm$^{-1}$.

Flux calibrations were not carried out because the IGRINS slit width is fixed. However, we derived the continuum fluxes of both H and K-band spectra based on the Two Micron All-Sky Survey (2MASS) magnitudes. The estimated H and K-band continuum fluxes are $6.40(\pm1.07)\times10^{-10}$ erg s$^{-1}$ cm$^{-2}$ μm$^{-1}$ (H = 9.80±0.20 mag, $A_H$ = 1.68 mag) and $4.38(\pm0.76)\times10^{-10}$ erg s$^{-1}$ cm$^{-2}$ μm$^{-1}$ ($K_s$ = 8.54±0.20 mag, $A_K$ = 1.08 mag) at $A_V$ = 9.6 mag (Dunham et al. 2013), respectively. The adopted extinction of $A_V$ = 9.6 mag is the derived mean value towards all Class II YSOs in Serpens star forming region based on optical studies.

We adopted 0.2 mag (around 17 percents in brightness) photometric uncertainties at both H and K-band magni- tudes due to near-infrared photometric variabilities toward YSOs in the Serpens, Orion A, ρ-Oph, ONC, IC348, and NGC 1333 (Kaas 1999; Carpenter et al. 2001; Scholz 2012). The maximum uncertainty in the edge of each band due to the assumption of the flux flatness inside each bandwidth is less than 10 percents by applying the same spectral index in each bandwidth as the derived spectral index of -1.27 between the measured H and K-band fluxes.

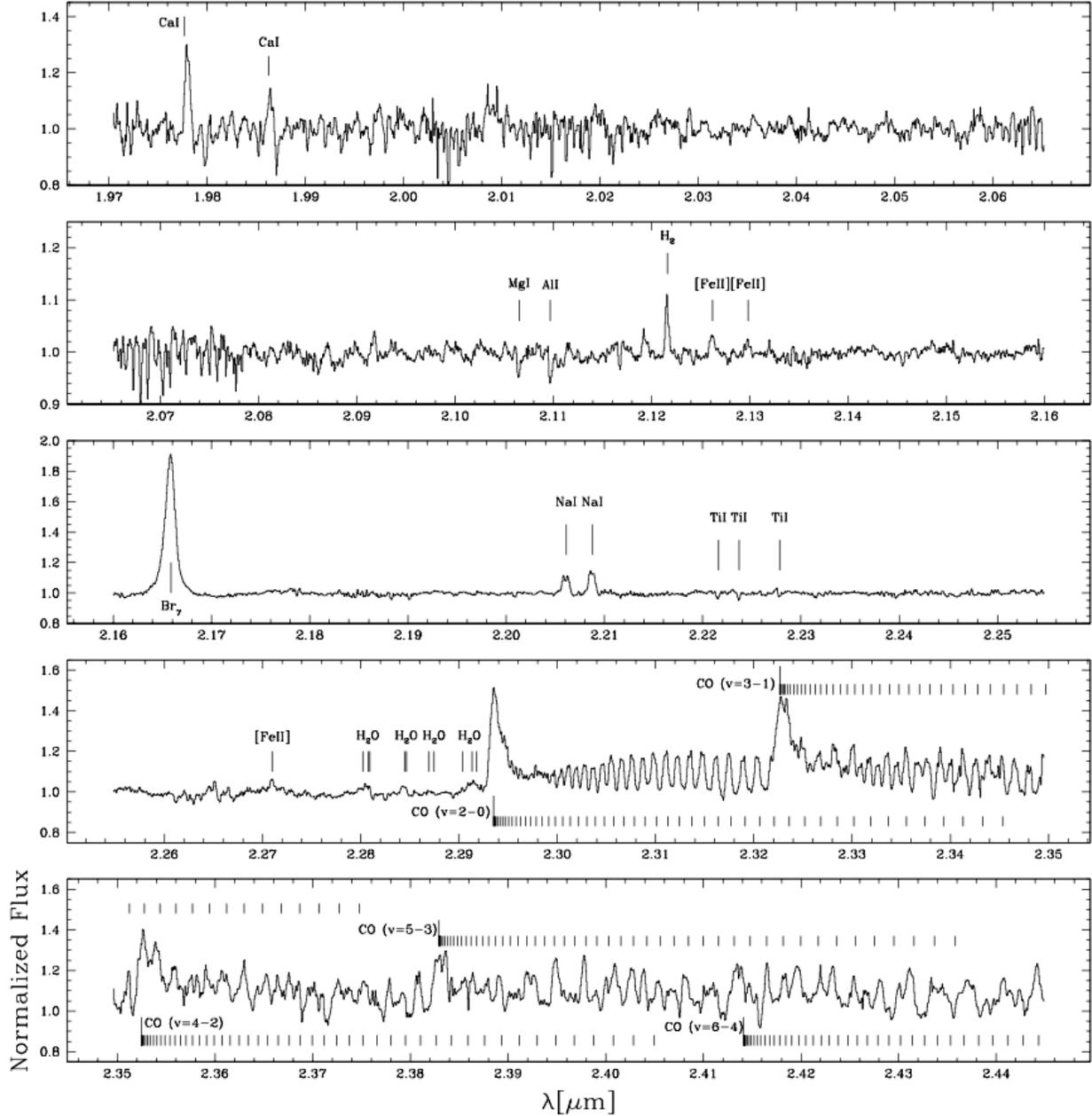

Figure 2. Telluric-corrected normalized IGRINS spectrum of the Flat-spectrum ESO Hα 279a covering the full $K$-band from 1.970 $\mu$m to 2.445 $\mu$m. Na I emission lines at 2.206, 2.209 $\mu$m and CO overtone emissions are clearly detected. Only $R$-branch lines of each CO overtone emission are marked. The normalized flux level is 4.38×10$^{-10}$ erg s$^{-1}$ cm$^{-2}$ $\mu$m$^{-1}$.

We also shifted the spectrum in wavelength with a total velocity shift of +22.33 km s$^{-1}$, which is measured from the comparison between the observed and the calculated wavelengths of the resolved CO ($v = 2 - 0$) R-branch lines. This shift is consistent with the expected shift using the $V_{helio}$=14.28 km s$^{-1}$ at the time of the observation (UT 2014-05-27, 10h 32m 57.939s) and the average $V_{LSR}$ of ~8 km s$^{-1}$ for the Serpens molecular clouds (McMullin et al. 2000; Lee et al. 2014).

With the IGRINS providing a large spectral grasp at high resolution for both the H and K-bands simultaneously, we detect many emission lines related with the inner disk activities, such as the Brackett line series from n = 19 – 4 to n = 7 – 4, the Ca I emission lines at 1.978 and 1.987 μm, the symmetric double-peaked Na I doublets at 2.206 and 2.209 μm, the $H_2$(v=1-0 S(1)) emission at 2.1218 μm, the $H_2O$ emission lines at the range of 2.28 μm to 2.29 μm, and the CO overtone line series from v = 2 – 0 to v = 6 – 4 (where v is the vibrational quantum number). In Table 1, we list all of these identified spectral lines except the unresolved $H_2O$ emission lines. We basically fit all the emission lines with Gaussian, Lorenzian, and Voigt profiles using the IRAF splot package to get the equivalent widths and center wavelengths. Weak emission lines are well fitted by the Voigt profile only. The hydrogen emission lines are quite well fitted by all three profiles. We estimated the error of the equivalent widths of the hydrogen lines from the different profile fittings. In Figure 2, we marked the ten strongest $H_2O$ emission lines from Table 2 in Thi & Bik (2005).

### *2.1. Identified stellar absorption and jet emission lines*

We identified some of stellar absorption lines, Fe I, Fe II, K I, Mg I, Al I, and Ti I, listed in Table 1. The center wavelength of the K I, Mg I, Al I, and Ti I lines match the expected wavelength, confirming the accuracy of the applied radial velocity.

We also detect ionized iron forbidden emission lines, [Fe II], [Fe III], and [Fe IV], which are known as jet tracers (Pyo et al. 2002) (see Table 1). The well-known jet tracers of [Fe II] emission lines at 1.533, 1.643, and 1.713 μm all show almost the same blue-shifted velocity of ~260 km s$^{-1}$. Overall, the derived ionized jet velocities of Fe forbidden emission lines are in the range of 200~260 km s$^{-1}$ and the average velocity line width is ~40 km s$^{-1}$. The true velocity range is 280~370 km s$^{-1}$ assuming the jet inclination of 45° (Aspin & Greene 2007). These high velocities, with narrow line widths, suggest that these detected ionized Fe forbidden emission lines trace the high-velocity component (HVC) of jets defined by Pyo et al. (2002). Particularly, the weakly detected 2.133 μm [Fe II] emission line draws our attention due to its highly blue-shifted velocity of 470 km s$^{-1}$, which is ~660 km s$^{-1}$ when we assume the jet inclination of 45°. If the inclination angle is correct, then this is an extremely high velocity, especially compared to the observed ~450 km s$^{-1}$ terminal velocity of HVC in low-mass YSOs suggested by Pyo et al. (2009). We need the high spatial and spectral resolution observation to investigate properties of this extremely high velocity jet component of 2.133 μm [Fe II].

### *2.2. $H_2$ emission at 2.1218 μm*

A molecular hydrogen $H_2$(v=1-0 S(1)) emission line at 2.1218 μm is detected. $H_2$ emission lines have been known as good tracers of the shocked regions in the molecular outflows (Wilking et al. 1990; Davis et al. 2008; Lim et al. 2012) as well as the accreting inner disks of YSOs (Bary et al. 2003, 2008). In our target source, $H_2$ emission line might be more closely related with the inner disk, since the measured radial velocity from the central wavelength of the emission is similar to the systemic velocity of the central source (see Table 1). We obtained a full width at half maximum (FWHM) of ~45 km s$^{-1}$ and a line width in the wing of ~100 km s$^{-1}$, respectively, from a Gaussian fitting. If the line broadening is only due to the disk kinematics, we find that $H_2$ emission is mainly from the disk in the range of around 0.3~1.3 AU assuming the central stellar mass of 1.5$M_\odot$ and the disk inclination of 45° (Aspin & Greene 2007). We, however, note that the emission line does not show the double-peaked profile which is the clear evidence of the rotational motion in the disk.

Table 1. Identified possible spectral lines towards ESO Hα 279a

| Species | $\lambda^{(1)}$ ($\mu$m) | $\lambda_{obs}^{(2)}$ ($\mu$m) | $\Delta v^{(3)}$ (km s$^{-1}$) | EW$^{(4)}$ (Å) | Species | $\lambda^{(1)}$ ($\mu$m) | $\lambda_{obs}^{(2)}$ ($\mu$m) | $\Delta v^{(3)}$ (km s$^{-1}$) | EW$^{(4)}$ (Å) |
|---|---|---|---|---|---|---|---|---|---|
| **H − band** | | | | | **K − band** | | | | |
| [Fe III] | 1.51458 | 1.51359 | −196 | −0.77 | Ca I | 1.97820 | 1.97816 | −6 | −1.90 |
| K I | 1.51631 | 1.51630 | −2 | 0.20 | Ca I | 1.98680 | 1.98674 | −9 | −0.65 |
| K I | 1.51684 | 1.51679 | −9 | 0.15 | Mg I | 2.10667 | 2.10668 | 1 | 0.12 |
| H I (Br 19) | 1.52606 | 1.52627 | 41 | −0.58±0.10 | Al I | 2.10988 | 2.10989 | 1 | 0.18 |
| [Fe II] | 1.53389 | 1.53252 | −268 | −0.30 | H$_2$ ($v$=1−0S(1)) | 2.12180 | 2.12178 | −3 | −0.38 |
| H I (Br 18) | 1.53418 | 1.53443 | 49 | −0.61±0.09 | [Fe II] | 2.12869 | 2.12641 | −321 | −0.26 |
| H I (Br 17) | 1.54389 | 1.54428 | 77 | −0.67±0.02 | [Fe II] | 2.13335 | 2.13000 | −470 | −0.15 |
| H I (Br 16) | 1.55565 | 1.55592 | 52 | −2.53±0.37 | H I (Brγ) | 2.16612 | 2.16601 | −15 | −16.39±0.33 |
| H I (Br 15) | 1.57007 | 1.57046 | 74 | −1.20±0.15 | Na I | 2.20625 | 2.20625 | 0 | −1.11 |
| [Fe III] | 1.57641 | 1.57512 | −245 | −0.53 | Na I | 2.20897 | 2.20892 | −7 | −1.23 |
| [Fe IV] | 1.57798 | 1.57683 | −218 | −0.70 | Ti I | 2.22173 | 2.22176 | 4 | 0.15 |
| H I (Br 14) | 1.58806 | 1.58839 | 62 | −3.59±0.87 | Ti I | 2.22389 | 2.22390 | 1 | 0.14 |
| Fe I | 1.58952 | 1.58941 | −21 | −0.54 | Ti I | 2.22801 | 2.22802 | 1 | 0.06 |
| H I (Br 13) | 1.61093 | 1.61132 | 73 | −2.50±0.34 | [Fe II] | 2.27273 | 2.27117 | −206 | −0.25 |
| H I (Br 12) | 1.64072 | 1.64108 | 66 | −3.12±0.57 | CO ($v=2-0$) | 2.29350 | − | − | − |
| [Fe II] | 1.64340 | 1.64257 | −260 | −2.23 | CO ($v=3-1$) | 2.32270 | − | − | − |
| Fe I | 1.67239 | 1.67237 | −4 | 0.27 | CO ($v=4-2$) | 2.35247 | − | − | − |
| Fe I | 1.67531 | 1.67552 | 38 | 0.55 | CO ($v=5-3$) | 2.38295 | − | − | − |
| Fe II | 1.67673 | 1.67679 | 11 | 0.23 | CO ($v=6-4$) | 2.41414 | − | − | − |
| H I (Br 11) | 1.68065 | 1.68101 | 64 | −5.09±0.79 | | | | | |
| [Fe III] | 1.69023 | 1.68942 | −143 | −1.21 | | | | | |
| [Fe II] | 1.71251 | 1.71111 | −245 | −0.39 | | | | | |
| [Fe III] | 1.73566 | 1.73417 | −257 | −0.23 | | | | | |
| H I (Br 10) | 1.73621 | 1.73663 | 73 | −4.90±0.71 | | | | | |

(1) $\lambda$ is one calculated from the difference between energy levels tabulated by NIST (National Institute of Standards and Technology) Atomic Spectra Database Lines Data. The wavelengths for forbidden lines are based on Peter van Hoof's Atomic Line List at http://www.pa.uky.edu/~peter/atomic/. The wavelengths of Ca I lines at $K$-band are from the Infrared Spectral Atlases of the Sun from NOAO (Wallace et al. 1996) (2) $\lambda_{obs}$ is the observed wavelength with the velocity correction of +22.33 km s$^{-1}$. (3) $\Delta v$ is the velocity difference between $\lambda$ and $\lambda_{obs}$. The minus sign represents the blue-shifted velocity compared to that of $\lambda$. (4) EW is the line equivalent width in Angstrom. The minus sign represents emission. Errors of the Brackett series are derived from three different equivalent measurements using the package of IRAF splot.

## 3. DISK ACCRETION RATE

Strong and sequential Brackett series of hydrogen recombination emission lines are observed in the H to K-band of ESO Hα 279a (see Fig. 1 and 2). Brackett emission series from n = 19 − 4 to n = 10 − 4 at H-band are all red-shifted compared to the rest velocity of the central star, while Brγ (n = 7–4) is slightly blue-shifted (see Table 1). The origin of the redshifted Brackett series is not well known.

However, the luminosity of Brγ emission line is strongly correlated with the disk accretion rate, and is thus useful for studying the accretion process in highly embedded YSOs at infrared wavelengths (Muzerolle et al. 1998b; Natta et al. 2004; Mendigutía et al. 2011). We calculate the Brγ luminosity with the observed equivalent width (-16.39 °A), assuming a K-band continuum flux of 4.38(±0.76)×10$^{-10}$ erg s$^{-1}$ cm$^{-2}$ $\mu$m$^{-1}$ at K$_s$ = 8.54±0.20 mag and A$_K$ = 1.08 mag derived by adopting A$_V$ = 9.6 mag (Dunham et al. 2013) and following the extinction law of Cardelli et al. (1989). The estimated Brγ luminosity found to be L$_{Brγ}$ = 1.78(±0.31)×10$^{31}$ erg s$^{-1}$ at the adopted distance of 429 pc. From the Brγ luminosity, we estimate the accretion luminosity L$_{acc}$ by the following relations; log(L$_{acc}$/L$_\odot$) = (1.26 ± 0.19) log(L$_{Brγ}$/L$_\odot$) + (4.43 ± 0.79) in Muzerolle et al. (1998b); log(L$_{acc}$/L$_\odot$) = 0.9[log(L$_{Brγ}$/L$_\odot$) + 4] − 0.7 in Natta et al.

(2004); $\log(L_{acc}/L_\odot) = (0.91 \pm 0.27) \log(L_{Br\gamma}/L_\odot) + (3.55 \pm 0.80)$ in Mendigutía et al. (2011). We can then estimate the mass accretion rate by $\dot{M}_{acc} = L_{acc} R_\star / GM_\star = 2 \sim 10 \times 10^{-7}\, M_\odot\, yr^{-1}$. The applied central stellar mass ($M_\star$) and radius ($R_\star$) for the calculation of mass accretion rate are 1.5 $M_\odot$ and 2.0 $R_\odot$, respectively.

Both mass (ranges of 1.5~2.2 $M_\odot$) and radius (ranges of 1.5~3.0 $R_\odot$) of the central star are derived from the evolutionary model of Siess et al. (2000), based on the magnitudes of V =20.6±0.20 mag and R=17.87±0.04 mag obtained by Aspin et al. (1994). The applied temperature and bolometric magnitude in the Hertzsprung-Russell (HR) diagram to obtain these stellar parameters are $T_{eff}$=6200±1500 K and $M_{bol}$=2.68±0.20 mag, respectively; the estimated color and absolute magnitude are $(V-R)_0$=0.33±0.24 mag and $M_{V_0}$=2.84±0.20 mag at the distance of 429 pc and $A_V$ = 9.6 mag. We adopt the bolometric correction and color-to-temperature relation of Kenyon & Hartmann (1995). We, however, note that the V magnitude could be affected by accretion process. The overestimated V luminosity compared to R luminosity might make this source to have higher mass and older age than the real.

As a result of this exercise, we find that the accretion properties of ~1.5 $M_\odot$ Flat-spectrum YSO are similar to those of other YSOs. For example, the obtained accretion rate of $2 \sim 10 \times 10^{-7}\, M_\odot\, yr^{-1}$ is consistent with the relations between the accretion rate and the central stellar mass, $\dot{M}_{acc} \propto M^{2.1}$ by Muzerolle et al. (2005) and $\dot{M}_{acc} \propto M^{1.8 \pm 0.2}$ by Natta et al. (2006), respectively.

## 4. MODELING OF THE CO OVERTONE EMISSION LINES

CO overtone emission lines are good tracers of the warm inner disks of YSOs (e.g. Najita et al. 2000, 2003; Wheel- wright et al. 2010; Cowley et al. 2012; Ilee et al. 2014). The most prominent features detected in a K-band spectrum of ESO Hα 279a are the CO overtone emission lines of v=2–0, v=3–1, v=4–2, v=5–3, and v=6–4. These features, shown in Figure 2, cover most wavelengths from 2.29 to 2.45 μm. Here, we apply the model fit from the first to fifth CO overtone emission lines, simultaneously, which provides more constraints on the determination of both temperature and density of warm CO gaseous inner disk compared to a model fitting only to the first CO overtone emission lines.

In our model, we assume a simple geometrically flat Keplerian rotating disk with a power law distribution in density, $N(r) = N_{in}(r/r_{in})^{-p}$, and temperature, $T(r) = T_{in}(r/r_{in})^{-q}$, where $r_{in}$ is the inner radius of the disk, $N_{in}$ and $T_{in}$ are the column density and temperature of the disk at the inner radius, respectively, and p and q are the power indices of the density and temperature distributions, respectively. We also assume the CO gas to be in local thermodynamic equilibrium (LTE) and consider the vibrational and rotational transitions up to v = 15 and J = 110 to calculate the total partition function of a CO molecule, where J is the rotational quantum number (see details of the modeling of the CO overtone emission lines in Kraus et al. 2000). The derived CO line profiles are smoothed with a Gaussian profile with the width of the IGRINS instrumental resolution of 7 km s$^{-1}$ at K-band. Since the estimated velocity broadening is only ~1 km s$^{-1}$ at the temperature of 4000 K, we do not consider the line broadening of CO gas due to its thermal motion.

There are six model parameters: p, q, $N_{in}$, $T_{in}$, $V_{in} \sin i$ and $V_{out} \sin i$, where $V_{in}$ and $V_{out}$ are the Keplerian velocities at the inner and outer radii and i is the inclination of the Keplerian orbit. To reduce the number of free parameters, we fix the power indices of the radial distributions of temperature and density as p = 1.5 and q = 0.5 following Hayashi (1981), respectively. For example, p = 1.5 and q = 0.4 are the typically obtained power indices in disks around YSOs (Guilloteau & Dutrey 1994; Guilloteau et al. 2011).

We define a goodness of fit,

$$\chi^2 = [F_{obs}(\lambda_i) - F_{mod}(\lambda_i)]^2, \quad (1)$$

where $F_{obs}$ and $F_{mod}$ are observed and calculated normalized fluxes, respectively. In order to calculate $\chi^2$ properly, we calculate model flux at each observed wavelength.

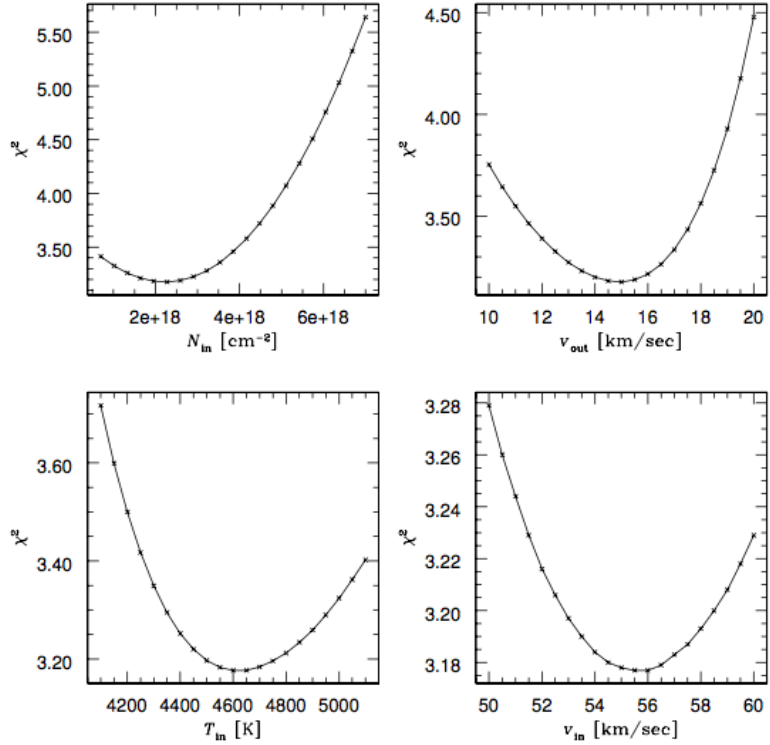

Figure 3. $\chi^2$ defined in equation (1) as a function of each of four free parameters. The four free parameters are $N_{in}$, column density at the inner disk edge, $T_{in}$, temperature in the inner disk edge, $V_{in} \sin i$, line-of-sight velocity at the inner disk edge, and $V_{out} \sin i$, line-of-sight velocity at the outer disk edge.

the four-dimensional parameter space, $N_{in}, T_{in}, V_{in} \sin i$ and $V_{out} \sin i$, we find a set of best fitting parameters which minimize $\chi^2$ values. The $\chi^2$ fitting results of four free parameters are shown in Figure 3 with the best fitting values of $N_{in} = 2.275 \times 10^{18}$ cm$^{-2}$, $T_{in} = 4600$ K, $V_{in} \sin i = 55.5$ km s$^{-1}$, and $V_{out} \sin i = 15.0$ km s$^{-1}$. Figure 4 shows the best fitting result with a red solid line and the observed CO overtone emission lines with a black solid line. In this Keplerian disk model, the derived temperature and column density mainly rely on the relative intensities among five CO overtone bandheads, while the derived velocities at the inner and outer radii mainly depend on the line width in the wing (maximum line width) and the separation between two peaks in the symmetric double-peaked line profiles of CO (v = 2 − 0) R-branch lines, respectively. In Figure 5, we overlap eleven CO (v = 2 − 0) R-branch lines from R(25) to R(15) in the range of wavelengths of 2.30 − 2.32 μm to show the consistency of these line features. The averaged line (solid line) shows the symmetric double-peaked profile, which is the characteristic feature of a Keplerian rotating disk.

The derived model provides a good fit through all five CO overtone emission bands, except in the first CO (v = 2–0) bandhead. We expect that the higher observed flux intensity at the first CO overtone bandhead compared to the model flux might be due to non-LTE effect, which we will explore in a separate paper. The mismatch at the bandhead of CO v = 2 − 0 does not significantly affect each line profile related with kinematical properties, such as velocities at the inner and outer radii of the CO disk. When we adopt the central stellar mass of 1.5 $M_\odot$ and assume 45° for the inclination of the CO disk, we find that CO gas emission is mainly from a Keplerian warm disk in the radial range from 0.22 AU to 3.00 AU, and $T_{in} = 4600$ K.

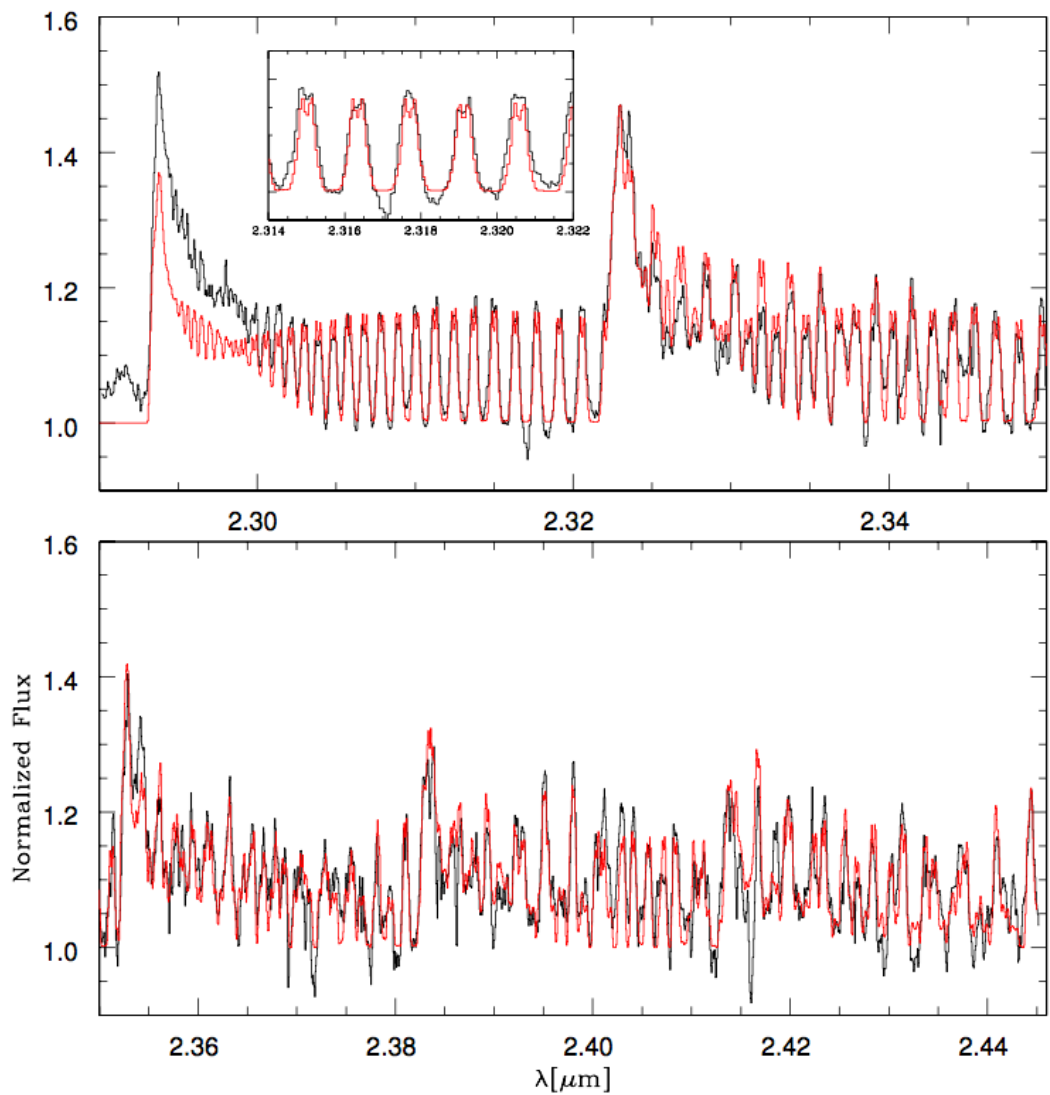

Figure 4. Comparison of observed CO overtone lines and calculated ones. The observed spectrum is shown with a black solid line and the calculated one is shown with a red line. The enlarged spectrum in the range of 2.314 - 2.322$\mu$m is shown in a small box of the upper panel. Some of lines in the box show a clear two-peaked profile, which is the evidence of the Keplerian rotating disk (see also Fig 5).

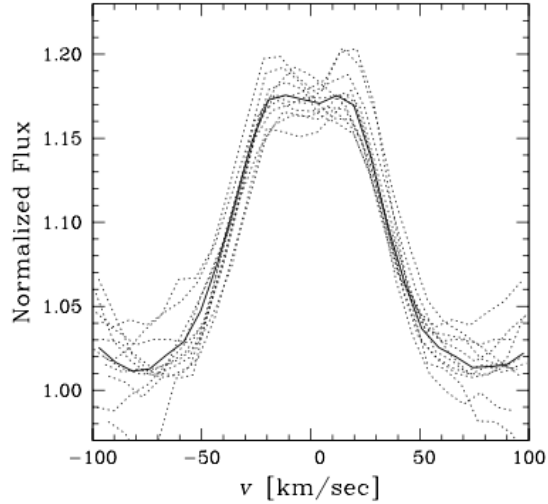

Figure 5. Dotted lines are 11 $R$-branch lines from $R(25)$ to $R(15)$ of CO ($v = 2 - 0$) in the range of wavelengths of 2.30784 − 2.32059 μm. A solid line is the averaged line profile of all 11 lines. It shows a symmetric double-peaked profile, which is an evidence of the Keplerian disk.

## 5. MODELING OF NA I AND CA I EMISSION LINES

We detect the rare Na I emission lines at both 2.206 and 2.209 μm and both lines display a symmetric double-peaked profile, especially the 2.206 μm Na I, which is the strong evidence of the Keplerian rotational disk (Fig. 6). There are asymmetries at the wings of both Na I emission lines showing that emission at the blue wing is broader than that at the red wing. Emission larger than 100 km s$^{-1}$ is more likely from the accretion flow and the asymmetry may be due to the redshifted absorption of infalling material (Muzerolle et al. 1998a). Assuming the optically thin and the LTE conditions, we expect that the ratio of two Na I lines at 2.206 μm and 2.209 μm is 2, which is in fact the ratio of statistical weights of $^2P_{3/2}$ and $^2P_{1/2}$ states. The observed intensity ratio is around 1, which means that Na I emitting gas is neither optically thin nor under LTE conditions. However we can still extract the kinematic properties of the Na I gaseous disk from both 2.206 and 2.209 μm Na I emission lines, even though 2.209 μm Na I emission line does not show the ideal double-peaked profile of the Keplerian rotating disk.

In order to model the Na I lines, we have first assumed a simple geometrically flat Keplerian disk with a power law distribution in density. We have only considered three parameters for fitting this symmetric and double-peaked profile; the line-of-sight velocity at the outer boundary of disk, $V_{out} \sin i$, the ratio of the inner to the outer radii, $r_{in}/r_{out}$, and the power index of the radial column density p. The power index of the density profile is limited to the parameter space between 1.0 and 2.5 with 0.5 step size, since the typical and most frequently used value is 1.5 (Guilloteau & Dutrey 1994; Guilloteau et al. 2011). Again, we smooth a calculated model line profile with a Gaussian with the width of the 7 km s$^{-1}$ IGRINS instrumental resolution. We do not consider the line broadening of Na I gas due to its thermal motion, since the estimated velocity broadening is only ~1.2 km s$^{-1}$ at the average temperature of 4000 K. We also use $\chi^2$ defined in equation (1) and find a set of parameters which minimize the $\chi^2$ values.

Figure 6(a) compares the observed line profile at 2.206 μm (a solid line) with two calculated model profiles for p = 1.5 (a dashed line) and p = 2.0 (a dotted line), respectively. The best fitting parameters are $r_{in}/r_{out}$ = 0.03, and $V_{out} \sin i$ = 30.53 km s$^{-1}$ for p = 1.5, and $r_{in}/r_{out}$ = 0.08, and $V_{out} \sin i$ = 28.42 km

s$^{-1}$ for p = 2.0, respectively. Figure 6(b) compares the observed line profile at 2.209 μm with a calculated model profile with p = 2.0. The best fitting parameters are $r_{in}/r_{out}$ = 0.03, and $V_{out}$ sini = 22.11 km s$^{-1}$. In this Keplerian disk model, the derived velocities at the inner and outer radii mainly rely on the line width in the wing (maximum line width) and the separation between two peaks in the symmetric double-peaked line profiles, respectively.

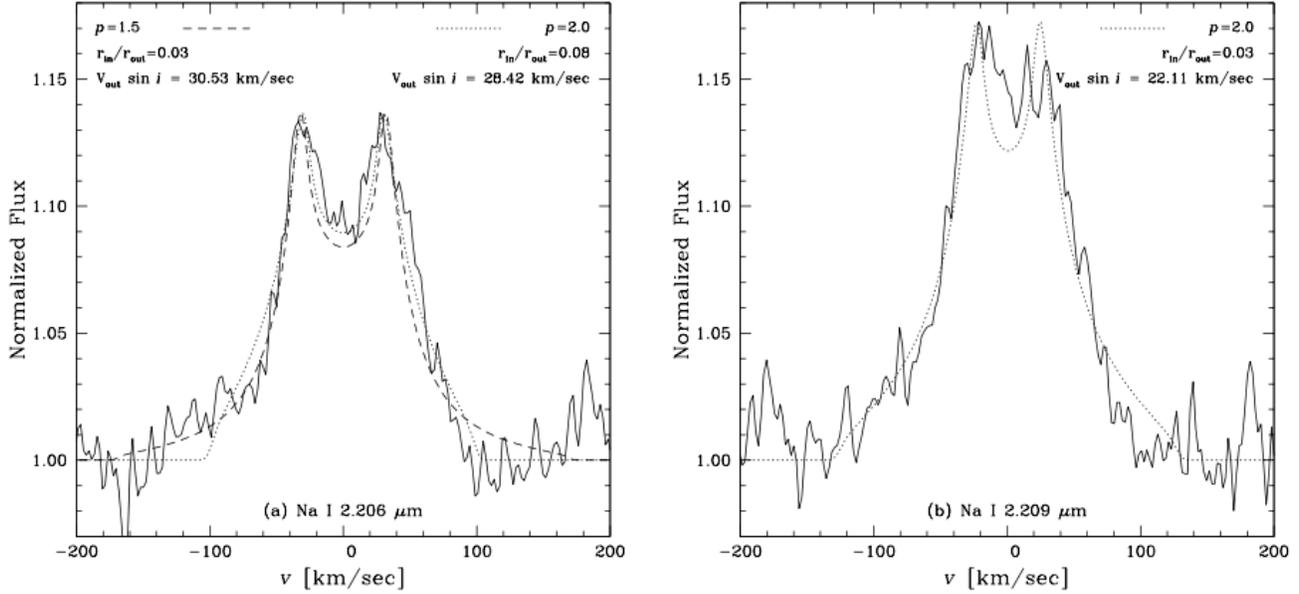

Figure 6. Na I emission lines at (a) 2.206 μm and (b) 2.209 μm, respectively. The line profiles calculated based on the Keplerian disks with the parameters written in each panel. In each panel, the observed spectrum is plotted with a solid line and model spectra are plotted with dotted or dashed lines.

We conclude that the 2.206 μm Na I emission is mainly from the inner disk region in the radial range from 0.02 AU to 0.72 AU with p=1.5 or from 0.07 AU to 0.83 AU with p=2.0, when we adopt the central stellar mass of 1.5 M$_\odot$ and the disk inclination of 45°. Most of the 2.209 μm Na I emission seems to come from the narrow inner hot disk region at the range from 0.04 AU to 1.36 AU with p=2.0.

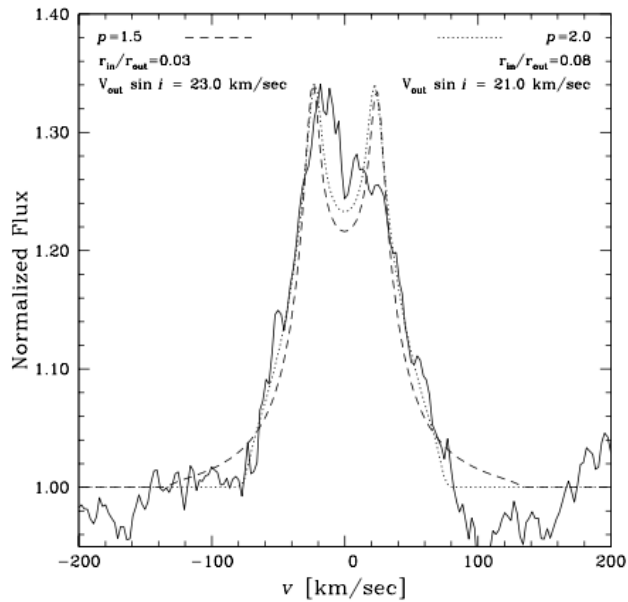

Figure 7. Ca I emission line at 1.978 μm. The line profiles calculated based on the Keplerian disks with the parameters written inside the panel. The observed spectrum is plotted with a solid line and model spectra are plotted with dotted or dashed lines.

Overall, Na I emissions mainly originate in the Keplerian disk with the range of 0.04~1.00 AU, where hot Jupiters reside. Confirming the Keplerian motion in this range of disk using Na I emission lines is important first step to understand how the inward migration stops and how hot Jupiters maintain their stable orbits.

Emission lines from Ca I at 1.978 and 1.987 μm are also detected. Davis et al. (2011) suggested that Ca I emission lines are associated with an accreting disk from the spectroscopic studies of seven embedded YSOs. In Figure 7, we show the 1.978 μm Ca I emission line (a solid line) with two Keplerian model profiles for p = 1.5 (a dashed line) and p = 2.0 (a dotted line), respectively. Overall, these two simple models provide good fits to the line profile, even though the line profile is not the perfect symmetric double-peaked feature. The best fitting parameters are $r_{in}/r_{out}$ = 0.03, and $V_{out}$ sini = 23.0 km s$^{-1}$ for p = 1.5, and $r_{in}/r_{out}$ = 0.08, and $V_{out}$ sini = 21.0 km s$^{-1}$ for p = 2.0, respectively. We find that 1.978 μm Ca I emission is mainly from the range of 0.04~1.25 AU at p=1.5 or 0.12~1.50 AU at p=2.0, assuming the disk inclination of 45°. We do not analyze the 1.987 μm Ca I emission line here, since the red-shifted part of the line is heavily contaminated by strong telluric lines.

## 6. DISCUSSIONS

Many people have fit LTE models to the bandhead of the first CO overtone series to get the parameters of gas disks. A partial list of examples is Kraus et al. (2000) for MWC 349, Aspin & Greene (2007) for ESO Hα 279a, Wheelwright et al. (2010) for seven intermediate/massive YSOs, and Ilee et al. (2014) for five Herbig Ae/Be stars observed with Very Large Telescope(VLT)/Cryogenic Infrared Echelle Spectrograph (CRIRES). In this paper, we are able to better determine the disk properties of ESO Hα 279a by the LTE modeling of all five CO overtone series from v = 2 − 0 to v = 6 − 4 covering the wavelength from 2.29 μm to 2.45 μm. Our model quite successfully fits the observed series except for the bandhead of the first overtone series. Near the bandhead of the first overtone series, the observed flux is larger than the flux calculated from the model. Furthermore, the observed peak flux of the bandhead of the first overtone is larger than that of the second overtone series (see Fig 4), which is the case where CO line opacity is almost infinite (Kraus et al. 2000). So we think that the bandhead of the first overtone series is more influenced by non-LTE effects than other parts of the spectrum.

In Figure 8, we show an example of the model fitting only for the first CO overtone bandhead. The best fitting parameters are $N_{in}$ = 2.575 × 10$^{19}$ cm$^{-2}$, $T_{in}$ = 5800 K, $V_{in}$ sin i = 56.0 km s$^{-1}$, and $V_{out}$ sin i = 10.0 km s$^{-1}$, respectively. We could nicely fit the observed first overtone bandhead spectrum with the parameters. However, we note that this model gives too much fluxes for the other CO bandheads than the observed one. Apparently, the derived values of temperature and density are too high for the rest of CO overtone bandheads. The velocity parameters determined from only the first overtone bandhead are not much different from those determined from all five CO overtone series. This is due to the fact that the kinematics is mostly determined by the widths of lines.

We have synthesized Na I emission lines with a Keplerian disk model and found that Na I emission lines are good tracers of the innermost disk. Na I emission lines have been detected towards only several Class I/Flat-spectrum YSOs, HH100IR (TS 2.6), RCrA IRS2 (TS 13.1) (Nisini et al. 2005; Doppmann et al. 2005), ESO Hα 279a (Aspin & Greene 2007), ISO 159, SVS 20N (Eiroa & Djupvik 2008), and IRAS 03220+3035N (Connelley & Greene 2010). Two YSOs among them, HH100IR (TS 2.6) and ESO Hα 279a, show double-peaked Na I emission lines. We carefully searched the above literature of all YSOs with Na I emission lines and found that all of them have CO overtone emission lines at 2.3 μm. However, not all of the YSOs emitting CO overtone lines show Na I lines. It seems that Na I emission lines trace the innermost disk region where accretion starts. Our model results of both Na I and CO lines show that Na I lines originate closer to the central star than the CO overtone lines. We find that the radius range of Na gaseous disk, 0.04~1.00 AU, where hot Jupiters reside, which supports the idea that the inner gas disk might control the inward migration processes of exoplanets (Lin et al. 1996). Furthermore, the derived average inner radius of the Na gaseous disk, $r_{in}$ = 0.04 AU (see section 6), is close to the typical corotation radius, 0.03 ~ 0.08 AU (Najita et al. 2003), where material accretes to the central star following magnetic field lines connecting between the star and disk in magnetospheric accretion models (Königl 1991; Shu et al. 1994; Calvet & Hartmann 1992).

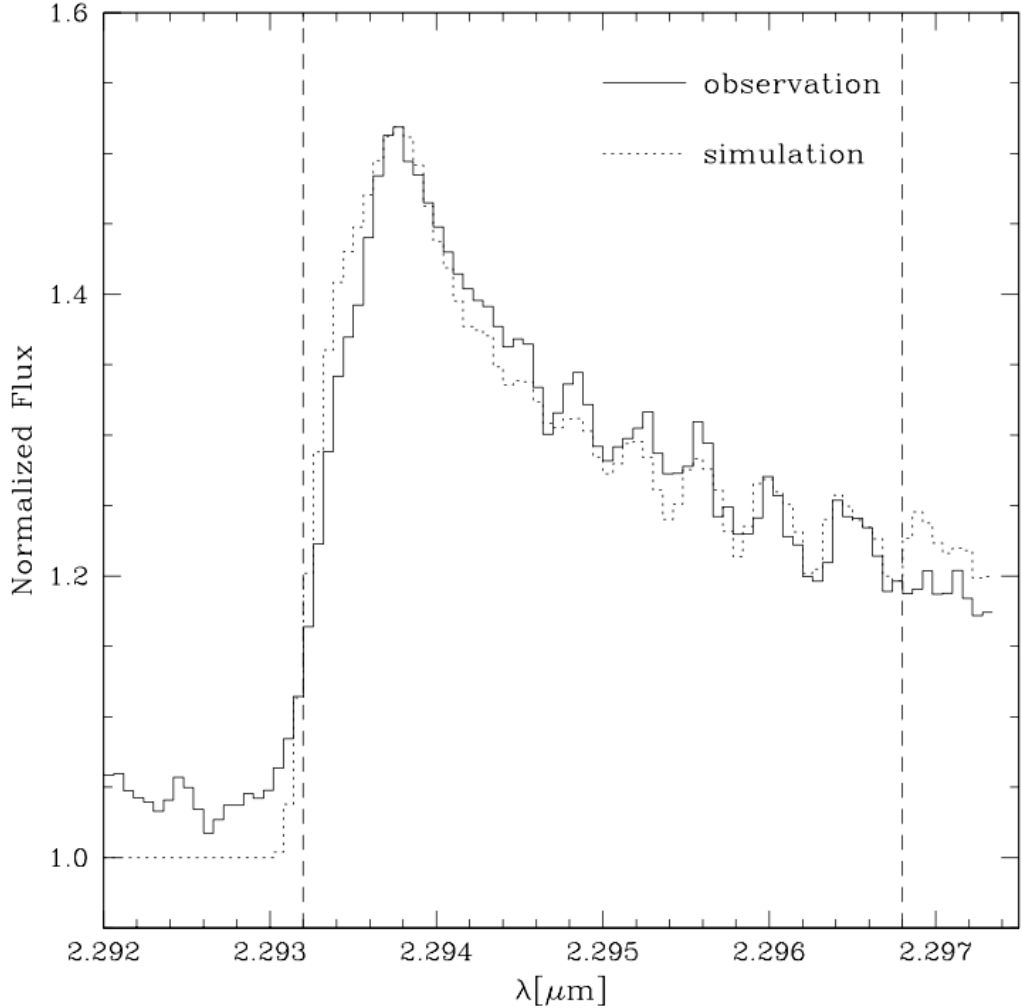

Figure 8. Comparison of the observed CO first overtone bandhead and calculated ones. The observed spectrum is shown with a solid line and the calculated one is shown with a dotted line. Two vertical dashed lines define the range for the calculation of $\chi^2$ to find the best fitting model. The best fitting parameters are $N_{in} = 2.575 \times 10^{19}$ cm$^{-2}$, $T_{in} = 5800$ K, $V_{in} \sin i = 56.0$ km s$^{-1}$, and $V_{out} \sin i = 10.0$ km s$^{-1}$.

However, there is a possibility that Na I symmetric double-peaked line profile might be due to the simple overlap of the stellar absorption line on top of the broad emission line, suggested by Aspin & Greene (2007). In fact, there are detections of stellar absorption lines in our observation such as Fe I, Fe II, K I, Mg I, Al I, and Ti I (see Fig. 1 and 2). In this interpretation, we suggest that the broad emission lines of Na I might be still related with disk, either from the farther outer disk region or from the disk wind region. We, however, do not expect that these emission lines are related with the jet/outflow or the infalling flow, because the central velocity of these emission lines are not blue- nor red-shifted (see Table 1).

Additionally, the dip in the center of the 1.978 μm Ca I emission can be also interpreted as the overlap of the stellar absorption line on top of the broad emission line. The central velocity of this emission line is also close to that of stellar radial velocity, which supports the disk origin of this gas emission.

## 7. SUMMARY

The near-infrared spectrometer IGRINS has the broadest spectral grasp at high-resolution and a single spectrum allows for the detailed study of the inner disk of YSOs at 1 AU radius. We have presented H and K-band spectra of a Flat-spectrum YSO, ESO Hα 279a, which show many atomic and molecular emission lines as well as the stellar absorption lines (see Fig. 1 and 2). We measure mass accretion rate of our target source with the Brγ emission line, and study the disk properties with CO overtone and Na I/Ca I emission lines.

Firstly, we derive the mass accretion rate of $2 \sim 10 \times 10^{-7}\,M_\odot\,\mathrm{yr}^{-1}$ from the measured Brγ luminosity of $1.78(\pm0.31)\times10^{31}\,\mathrm{erg\,s^{-1}}$ at the distance of 429 pc. We find that the accretion property of the 1.5 $M_\odot$ Flat-spectrum ESO Hα 279a is similar to those of other YSOs (Muzerolle et al. 1998b; Natta et al. 2004; Mendigutía et al. 2011).

Secondly, we find the inner warm rotating CO gaseous disk in the radial range from 0.22 AU to 3.00 AU by modeling the five CO overtone series of v=2–0, v=3–1, v=4–2, v=5–3, and v=6–4. In this model, we assume a simple geometrically flat Keplerian rotational disk with the power indices, p = 1.5 and q = 0.5, for density and temperature distributions, respectively, and the disk inclination of 45°. We find the disk properties of $N_{in} = 2.275 \times 10^{18}\,\mathrm{cm}^{-2}$, $T_{in} = 4600$ K, $V_{in} \sin i = 55.5\,\mathrm{km\,s^{-1}}$, and $V_{out} \sin i = 15.0\,\mathrm{km\,s^{-1}}$, which are determined by best-fitting model parameters.

Lastly, we find the Na I gaseous Keplerian disk in the radial range of 0.02 AU ~ 0.72 AU at p=1.5 and 0.07 AU ~ 0.83 AU at p=2.0 by modeling the symmetric double-peaked 2.206 μm Na I emission line. In the case of 2.209 μm Na I line, this emission is mainly from the radial range of 0.04 AU ~ 1.36 AU at p=2.0. We also find that Ca I emission originates in the similar disk region as Na I, based on the Keplerian model to 1.978 μm Ca I spectral line.

This work used the Immersion Grating Infrared Spectrograph (IGRINS) that was developed under a collaboration between the University of Texas at Austin and the Korea Astronomy and Space Science Institute (KASI) with the financial support of the US National Science Foundation under grant ASTR1229522, of the University of Texas at Austin, and of the Korean GMT Project of KASI. The IGRINS pipeline package PLP was developed by Dr. Jae-Joon Lee at Korea Astronomy and Space Science Institute and Professor Soojong Pak's team at Kyung Hee University.